\documentclass[paper,11pt]{paper}
\pdfoutput=1
\usepackage[T1]{fontenc}
\usepackage{mdframed}
\usepackage{comment}
\usepackage{mathtools}
\usepackage{upgreek}
\usepackage{graphicx}
\usepackage{amsmath}
\usepackage{amsfonts}
\usepackage{amssymb}
\usepackage{bm}
\usepackage{amsthm}
\usepackage{slashed}
\usepackage{enumerate,enumitem}
\usepackage{amsbsy}
\usepackage{bbm}
\usepackage{mathrsfs}
\usepackage{url}
\usepackage{bbm}
\usepackage{color}
\usepackage{framed}
\usepackage{subfig}
\usepackage{caption}
\usepackage{float}
\usepackage{cancel}
\usepackage{epstopdf}
\usepackage{jheppub}
\usepackage{hyperref}

\def\({\left(}
\def\){\right)}
\def\[{\left[}
\def\]{\right]}
\def\one{{\rm 1\kern -.9mm l}}              

\def\beq{\begin{equation}}
\def\eeq{\end{equation}}
\def\beqa{\begin{eqnarray}}
\def\eeqa{\end{eqnarray}}
\newcommand{\eqa}{\begin{eqnarray}}
\newcommand{\ena}{\end{eqnarray}}

\newcommand{\cA}{\mathcal{A}}
\newcommand{\cJ}{\mathcal{J}}
\newcommand{\cO}{\mathcal{O}}
\newcommand{\cH}{\mathcal{H}}
\newcommand{\cP}{\mathcal{P}}

\newcommand{\cL}{\mathcal{L}}

\newcommand{\cS}{\mathcal{S}}

\newcommand{\Bx}{{\bf x}}
\newcommand{\By}{{\bf y}}

\newcommand{\e}{\varepsilon}
\DeclareMathAlphabet{\mathpzc}{OT1}{pzc}{m}{it}

\DeclareMathOperator{\sech}{sech}

\newcommand{\R}{{\mathbb{R}}}

\newcommand\blank[1]{}

\newcommand{\fract}[2]{{\textstyle\frac{#1}{#2}}}

\newcommand{\balpha}{\alpha\kern -6.7pt\alpha}
\newcommand{\bbalpha}{\alpha\kern-4.95pt\alpha}

\newcommand\en{\end{equation}}
\newcommand\bea{\begin{eqnarray}}
\newcommand\eea{\end{eqnarray}}

\newcommand{\One}{{\hbox{{\rm 1{\hbox to 1.5pt{\hss\rm1}}}}}}
\renewcommand{\One}{{\mathbb 1}}
\renewcommand{\One}{{\rm 1\!\!1}}

\newcommand{\ba}{\begin{eqnarray}}
\newcommand{\ea}{\end{eqnarray}}

\newcommand{\be}{\begin{equation}}
\newcommand{\ee}{\end{equation}}

\newcommand\by{{\bf  y }}

\newcommand{\TbT}{\text{T}\bar{\text{T}}}

\newcommand{\JbT}{\text{J}\bar{\text{T}}}

\newcommand{\ta}{\tau}

\renewcommand{\ell}{{\mathcal L}}

\setlist[itemize]{leftmargin=*}
\newcommand{\NR}{\text{\tiny NR}}
\newcommand{\NLS}{\text{\tiny NLS}}
\newcommand{\shG}{\text{\tiny sh-G}}
\newcommand{\MF}{\text{\tiny MF}}
\definecolor{purple_nice}{rgb}{0.4,0.2,0.7}
\definecolor{fuel_blue}{RGB}{42,162,185}
\definecolor{YInMn_blue}{RGB}{46, 80, 144}
\definecolor{ultramarine}{RGB}{63, 0, 255}
\definecolor{KLEIN_blue}{rgb}{0, 0.18, 0.65}

\def\XXint#1#2#3{{\setbox0=\hbox{$#1{#2#3}{\int}$ }
\vcenter{\hbox{$#2#3$ }}\kern-.6\wd0}}

\title{\boldmath $\text{T}\bar{\text{T}}$-deformed Nonlinear Schr\"odinger }

\author[1]{Paolo Ceschin,}
\author[2]{Riccardo Conti,}
\author[1]{Roberto Tateo}

\affiliation[1]{Dipartimento\ di Fisica and Arnold-Regge Center, Universit\`a di Torino, and INFN, Sezione di Torino, Via P. Giuria 1, I-10125 Torino, Italy.}
\affiliation[2]{Grupo de F\'isica Matem\'atica da Universidade de Lisboa, Av. Prof. Gama Pinto 2, 1649-003 Lisboa, Portugal.}

\emailAdd{paolo.ceschin@edu.unito.it}
\emailAdd{rconti@fc.ul.pt}
\emailAdd{roberto.tateo@unito.it}
\abstract{ 
The $\text{T}\bar{\text{T}}$-deformed classical Lagrangian of a 2D Lorentz invariant theory can be derived from the original one, perturbed only at first order by the bare  $\text{T}\bar{\text{T}}$ composite field, through a field-dependent change of coordinates.  Considering, as an example, the nonlinear Schr\"odinger (NLS) model with generic potential, we apply this idea to non-relativistic models. The form of the deformed Lagrangian contains a square-root and is similar but different from that for relativistic bosons. We study the deformed bright, grey and Peregrine's soliton solutions. Contrary to naive expectations, the $\text{T}\bar{\text{T}}$-perturbation of nonlinear Schr\"odinger NLS with quartic potential does not trivially emerge from a standard non-relativistic limit of the deformed sinh-Gordon field theory. The  $c \rightarrow \infty$ outcome corresponds to a different type of irrelevant deformation. We derive the corresponding Poisson bracket structure, the equations of motion and discuss various interesting aspects of this alternative type of  perturbation, including  links with the recent  literature. }
\begin{document}
\maketitle
\flushbottom
\section{Introduction}
The presence of an irrelevant operator in a quantum field theory is
usually not good news, as far as understanding the high-energy physics of the model is concerned. Trying to reverse the renormalisation group flow requires the introduction of an infinite number of counterterms in the Lagrangian. Therefore, perturbing a field theory with irrelevant operators can drastically affect the ultraviolet properties of the model and introduce new fundamental degrees of freedom at high-energy.
In two space-time dimensions,  the $\TbT$ composite operator \cite{Zamolodchikov:2004ce} is an exception to this rule since this irrelevant field is well defined also at the quantum level.
The $\TbT$ perturbation is solvable \cite{Smirnov:2016lqw, Cavaglia:2016oda}, in the sense that physical observables of interest, such as the  S-matrix and the finite-volume spectrum, can be found in terms of the corresponding undeformed quantities. For the $\TbT$  operator, we can reverse the
renormalisation group trajectory and gain exact information about
ultraviolet physics. The outcome is stunning:  while the low-energy physics
resembles that of a conventional local quantum field theory  at
high-energy the density of states on a cylinder shows Hagedorn growth
similar to that of a string theory \cite{Dubovsky:2012wk, dubovsky2012effective,Caselle:2013dra}. 

A widely studied Lorentz-breaking perturbation is  the $\JbT$ model \cite{Guica:2017lia},  other integrable deformations that explicitly break this symmetry were introduced and partially studied in \cite{Conti:2019dxg}. A framework where $\TbT$-type perturbations may potentially lead to concrete applications in fluid dynamics, nonlinear optics and  condensed matter physics corresponds to the domain of non-relativistic nonlinear wave equations.

One of the most-studied models with direct relevance in cold atom experiments is the nonlinear Schr\"odinger (NLS) equation. The primary purpose of this paper is to derive the explicit form of the $\TbT$ perturbed Lagrangian for a family of NLS equations with arbitrary interacting potential. The exact expression of the Lagrangian is surprisingly similar to that of $\TbT$-deformed relativistic bosons \cite{Cavaglia:2016oda, Bonelli:2018kik, Conti:2018jho}.  A second type of deformation of the  NLS model is obtained performing the non-relativistic limit of the deformed sinh-Gordon theory.

Refs \cite{Cardy:2020olv,Jiang:2020nnb}  appeared as we were working on this project. These authors discuss various aspects of deformed 1D non-relativistic quantum particle models with complementary results, compared to those presented here.

\section{Non-relativistic scalar theory: generalities}
In this paper we shall consider a class of non-relativistic classical field theories of a complex scalar field $\psi(\Bx)$, with $\Bx=(t,x)$, described by the hermitian Lagrangian density
\begin{equation}
\label{eq:undefLag}
\cL(\Bx)=\cL_K(\Bx)-V \;,\quad \cL_K(\Bx) = \frac{i}{2}\left(\psi^*\psi_{,t}-\psi\psi_{,t}^*\right)-\frac{\psi_{,x}\psi_{,x}^*}{2m} \;,
\end{equation}
where $\psi^*$ denotes the complex conjugate field, $\psi_{,t}:=\partial_{t}\psi$ and $\psi_{,x}:=\partial_x\psi$ are the derivatives  w.r.t. time and space, and the potential $V$ is a generic function of $|\psi|^2=\psi\psi^*$. Setting $V=V_{\NLS}:=g|\psi|^4$, for some real constant $g$, \eqref{eq:undefLag} becomes the nonlinear Schr\"odinger (NLS) Lagrangian density, i.e. $\cL=\cL_{\NLS}$.\\
Let us recall some well-known facts about these non-relativistic Lagrangian models. The dynamics of the system is described by the Euler-Lagrange equations
\begin{equation}
\begin{cases}
\displaystyle{\partial_x\left(\frac{\partial\cL}{\partial\psi_{,x}}\right) + \partial_{t}\left(\frac{\partial\cL}{\partial\psi_{,t}}\right)= \frac{\partial\cL}{\partial\psi}} \vspace{0.2cm}\\
\displaystyle{\partial_x\left(\frac{\partial\cL}{\partial\psi_{,x}^*}\right) + \partial_{t}\left(\frac{\partial\cL}{\partial\psi_{,t}^*}\right)= \frac{\partial\cL}{\partial\psi^*}}
\end{cases} \longrightarrow\quad
\begin{cases}
\displaystyle{i\psi_{,t}^* = \frac{1}{2m}\psi_{,xx}^*-V'\psi^*}\vspace{0.2cm}\\
\displaystyle{-i\psi_{,t} = \frac{1}{2m}\psi_{,xx}-V'\psi}
\end{cases} \;,
\end{equation}
where $\displaystyle{V':=\frac{\partial V}{\partial|\psi|^2}}$. The invariance under space and time translations implies the existence of a Noether current that is the stress-energy tensor, whose components are computed from $\cL$ as
\begin{equation}
\label{eq:stressgen}
    \displaystyle{T^\mu_{\;\;\nu}=\frac{\partial\cL}{\partial\psi_{,\mu}}\psi_{,\nu}+\frac{\partial\cL}{\partial\psi^*_{,\mu}}\psi^*_{,\nu}-\delta^\mu_{\;\;\nu}\cL} \;,\quad \mu,\nu=\lbrace t,x \rbrace \;,
\end{equation}
explicitly,
\begin{align}
\label{eq:TNLS}
&T^t_{\;\,t}(\Bx) = \frac{\psi_{,x}\psi_{,x}^*}{2m} + V \;,\quad T^t_{\;x}(\Bx) = \frac{i}{2}(\psi^*\psi_{,x}-\psi\psi_{,x}^*) \;,\notag \\ 
&T^x_{\;\;t}(\Bx) = -\frac{1}{2m}(\psi_{,x}^*\psi_{,t}+\psi_{,x}\psi_{,t}^*) \;,\quad T^x_{\;\;x}(\Bx) = -\frac{i}{2}\left(\psi^*\psi_{,t}-\psi\psi_{,t}^*\right)-\frac{\psi_{,x}\psi_{,x}^*}{2m} + V \;.
\end{align}
Let us define the currents $\cH^\mu=T^\mu_{\;\;t}$ and $\cP^\mu=T^\mu_{\;\;x}$ that fulfil the continuity equations
\begin{equation}
\label{eq:conteq}
\partial_{\mu} \cH^\mu = \partial_{\mu} \cP^\mu=0 \;,
\end{equation}
where the quantities $\cH:=\cH^t$ and $\cP:=\cP^t$ represent the total energy and momentum densities. The invariance under a global phase rotation implies the existence of the Noether current $\cJ$ whose components are computed from $\cL$ as
\begin{equation}
\cJ^\mu = -im\left(\frac{\partial\cL}{\partial\psi_{,\mu}}\psi-\frac{\partial\cL}{\partial\psi^*_{,\mu}}\psi^*\right) \;,\quad \mu\in\{t,x\} \;,
\end{equation}
explicitly,
\begin{equation}
\cJ^t=m|\psi|^2\;,\quad \cJ^x = \frac{i}{2}\left(\psi\psi_{,x}^*-\psi^*\psi_{,x}\right) \;.
\end{equation}
The current $\cJ$ fulfil the continuity equation
\begin{equation}
\partial_\mu\cJ^\mu = 0 \;,
\end{equation}
and the quantity $\mathcal{M}:=\cJ^t$ defines the total mass density. In Hamiltonian formalism, we introduce the Hamiltonian density
\begin{equation}
\cH(\Bx)= \psi_{,t}\pi+\psi_{,t}^*\pi^*-\cL(\Bx) = \frac{\psi_{,x}\psi_{,x}^*}{2m} + V \;,
\end{equation}
where $\pi$ and $\pi^*$ are the conjugated momenta defined by the Legendre transformation
\begin{equation}
\pi=\frac{\partial\cL}{\partial\psi_{,t}} = \frac{i}{2}\psi^* \;,\quad \pi^*=\frac{\partial\cL}{\partial\psi_{,t}^*} = -\frac{i}{2}\psi \;.
\end{equation}
Clearly, it is not possible to express the time derivative in terms of the conjugated momenta as the usual Legendre procedure would require. In fact, the last equations reveal the presence of redundant variables that is a typical feature of constrained Hamiltonian systems. As it is well-known, the Dirac-Bergmann algorithm allows to elegantly overcome this issue (see, for example \cite{GERGELY2002394}). However, due to the mixing between space and time coordinates, the situation appears to be much more complicated in the $\TbT$-deformed model. In the following, we shall mainly ignore this problem and postpone a rigorous study of the deformed Hamiltonian and Poisson structure to the future.
\section{The deformed Lagrangian}
\label{DeformedL}
The aim of this section is to derive the Lagrangian density $\cL(\Bx,\ta)$ of the $\TbT-$deformed theory. By definition, the latter fulfils the flow equation
\begin{equation}
\label{eq:floweq}
\partial_\ta \cL(\Bx,\ta) = \det{\left[T(\Bx,\ta)\right]} \;,\quad \cL(\Bx,0)=\cL(\Bx) \;,
\end{equation}
where $T(\Bx,\ta)$ is the deformed stress-energy tensor that descends from formula \eqref{eq:stressgen} setting $\cL=\cL(\Bx,\ta)$. In principle, equation \eqref{eq:floweq} can be solved for $\cL(\Bx,\ta)$ by means of a perturbative expansion around $\ta=0$ \cite{Cavaglia:2016oda}. However, the form of the original stress-energy tensor \eqref{eq:TNLS} discourages the application of this approach.

It is natural to try to obtain the deformed Lagrangian using the same change of variables found in \cite{Conti:2018tca} in the relativistic context and then check if the flow equation is satisfied.\footnote{
In \cite{Frolov:2019nrr, Frolov:2019xzi}, a general light-cone gauge approach to $\TbT$  was developed which can be used to find the deformed Lagrangians also of non-relativistic models.
In  \cite{Frolov:2019nrr}, the author claims to have obtained the  $\TbT$-deformed NLS Lagrangian, without presenting explicitly the result. For this reason, we are currently unable to make a direct comparison with our outcome.}
Following the logic of \cite{Conti:2018tca, Conti:2019dxg} and \cite{Coleman:2019dvf}, the starting point is the identity
\begin{equation}
\label{eq:actiondef}
\cA(\tau)=\int dt\,dx \,\cL(\Bx,\tau) = \int dt'\,dx' \left (\cL(\By)-\tau\det{\left[T(\By)\right]}\right) \;,
\end{equation}
where $\By=(t',x')$ is another set of coordinates related to $\Bx=(t,x)$ via a coordinate transformation $\By(\Bx)$ whose Jacobian $J$ is such that 
\begin{equation}\label{eq:J-1}
J^{-1}(\By)=\left(
\begin{array}{cc}
\displaystyle{\frac{\partial t}{\partial t'}} & \displaystyle{\frac{\partial t}{\partial x'}}\vspace{0.1cm}\\
\displaystyle{\frac{\partial x}{\partial t'}} & \displaystyle{\frac{\partial x}{\partial x'}}
\end{array}\right) = \left(
\begin{array}{cc}
1+\tau\,T^x_{\;\;x}(\By) & -\tau\,T^t_{\;x}(\By) \\
 -\tau\,T^x_{\;\;t}(\By) & 1+\tau\,T^t_{\;\,t}(\By)
\end{array}
\right) \;.
\end{equation}
Performing the coordinate transformation in \eqref{eq:actiondef} leads to
\begin{equation}
\label{eq:lagdef}
\cL(\Bx,\tau) = \frac{\cL(\By(\Bx))-\tau\det{\left[T(\By(\Bx))\right]}}{\det{\left[J^{-1}(\By(\Bx))\right]}} \;,
\end{equation}
that allows  to reconstruct $\cL(\Bx,\tau)$ from the original Lagrangian and the knowledge of the coordinate transformation.\\

\noindent
{\it Remark.} The Hessian matrix associated to this change of variables is symmetric on-shell, in fact
\begin{gather}
\frac{\partial}{\partial x'}\frac{\partial t}{\partial t'} = \ta\partial_{x'}T^x_{\;\;x}(\By) = -\ta\partial_{t'}T^t_{\;x}(\By) = \frac{\partial}{\partial t'}\frac{\partial t}{\partial x'} \;,\notag \\
\frac{\partial}{\partial x'}\frac{\partial x}{\partial t'} = -\ta\partial_{x'}T^x_{\;\;t}(\By) = \ta\partial_{t'}T^t_{\;\,t}(\By) = \frac{\partial}{\partial t'}\frac{\partial x}{\partial x'} \;,
\end{gather}
due to the continuity equations \eqref{eq:conteq}.\\

\noindent
Let us sketch the computation of $\cL(\Bx,\tau)$ starting from formula \eqref{eq:lagdef}. The first step consists in writing the derivatives $\psi_{,t'}$, $\psi_{,x'}$ and their c.c. (complex conjugates) in terms of $\psi_{,t}$, $\psi_{,x}$ and their c.c. by inverting the algebraic systems
\begin{equation}
\left(\begin{array}{c}
\psi_{,t'} \\
\psi_{,x'}
\end{array}\right) = \left(J^{-1}(\By)\right)^{\text{T}}
\left(\begin{array}{c}
\psi_{,t} \\
\psi_{,x}
\end{array}\right) \;,\quad 
\left(\begin{array}{c}
\psi_{,t'}^* \\
\psi_{,x'}^*
\end{array}\right) = \left(J^{-1}(\By)\right)^{\text{T}}
\left(\begin{array}{c}
\psi_{,t}^* \\
\psi_{,x}^*
\end{array}\right) \;.
\end{equation}
A straightforward computation gives 
\begin{align}
\label{eq:dertransf}
\psi_{,t'} = \frac{2m\left(B-\cS\right)\psi_{,t}^*+2\tilde\tau A^*\left(\psi_{,x}^*\psi_{,t}-\psi_{,x}\psi_{,t}^*\right)}{2\tau (A^*)^2} \;,\quad \psi_{,x'} = \frac{2m\left(B-\cS\right)}{2\tau A^*} \;,\notag \\
    \psi^{*}_{,t'} = \frac{2m\left(B-\cS\right)\psi_{,t}-2\tilde\tau A\left(\psi_{,x}^*\psi_{,t}-\psi_{,x}\psi_{,t}^*\right)}{2\tau A^2} \;,\quad \psi^{*}_{,x'} = \frac{2m\left(B-\cS\right)}{2\tau A} \;,
\end{align}
where we defined
\begin{equation}
\label{eq:SBAdef}
\cS=\sqrt{B^2-\frac{2\tilde\tau}{m}AA^*} \;,\quad \tilde\tau=\tau(1+\tau V) \;,
\end{equation}
and
\begin{gather}
A=\psi_{,x}+\frac{i\tau}{2}\psi\left(\psi_{,x}^*\psi_{,t}-\psi_{,x}\psi_{,t}^*\right) \;,\quad A^*=\psi_{,x}^*+\frac{i\tau}{2}\psi^*\left(\psi_{,x}^*\psi_{,t}-\psi_{,x}\psi_{,t}^*\right) \notag \;,\\
B=1+\frac{i\tau}{2}\left(\psi^*\psi_{,t}-\psi\psi_{,t}^*\right) \;.
\end{gather}
Next, we write numerator and denominator of \eqref{eq:lagdef} in the following way
\begin{gather}
\det{\left[J^{-1}(\By)\right]} = \left(\frac{\tilde\tau}{\tau}-\frac{\tau}{2m}\psi_{,x'}\psi_{,x'}^*\right) - \tau\left(\cL(\By)-\tau\det{\left[T(\By)\right]}\right) \;, \notag \\
\cL(\By)-\tau\det{\left[T(\By)\right]} = \cL(\By)\left(\frac{\tilde\tau}{\tau}-\frac{\tau}{2m}\psi_{,x'}\psi_{,x'}^*\right) + \frac{i\tau}{4}\left(\psi_{,x'}^*\psi_{,t'}-\psi_{,x'}\psi_{,t'}^*\right)\left(\psi^*\psi_{,x'}-\psi\psi_{,x'}^*\right) \;.
\label{eq:comp1}
\end{gather}
Implementing the transformation \eqref{eq:dertransf} in the expressions above and after some algebraic manipulations one gets
\begin{multline}
\cL(\By(\Bx))-\tau\det{\left[T(\By(\Bx))\right]} = \frac{2i\,\tilde\tau^3}{m\tau^2(B-\cS)(B+\cS)^2}\Bigl(A^2\psi^*\psi^*_{,t}-(A^*)^2\psi\psi_{,t} + B\left(\psi^*_{,x}\psi_{,t}-\psi_{,x}\psi^*_{,t}\right) \\
\times\left(A\psi^*+A^*\psi\right)\Bigr) - \frac{i\tilde\tau^2(B-\cS)}{\tau^2(B+\cS)^2}\left(\psi^*\psi_{,t}-\psi\psi^*_{,t}\right) + \frac{2\tilde\tau\cS\left(\cS-B(1+2\tau V)\right)}{\tau^2(B+\cS)^2} \;.
\label{eq:comp2}
\end{multline}
Finally, plugging \eqref{eq:SBAdef} in \eqref{eq:comp2} and using the expressions \eqref{eq:comp1} in \eqref{eq:lagdef} one finds
\begin{equation}
\label{eq:lagdef2}
\boxed {\cL(\Bx,\tau) = -\frac{V}{1+\tau V} + \frac{-2+B + \cS}{2\tilde\tau} \,.}
\end{equation}
The form of the Lagrangian (\ref{eq:lagdef2}) is similar to its relativistic counterpart (see \cite{Cavaglia:2016oda, Conti:2018jho,Bonelli:2018kik}) and it would be nice to have some interpretation in terms of topological gravity \cite{Dubovsky:2018bmo}. As stated at the beginning of the section, we can easily check that \eqref{eq:lagdef2} fulfils \eqref{eq:floweq} using the following expressions
\begin{align}
&T^x_{\;\;x}(\Bx,\tau)=\frac{1}{\tau}-\frac{B\left(B+\cS\right)}{2\tilde\tau\,\cS} \;,\quad T^x_{\;\;t}(\Bx,\tau)=-\frac{1}{2m\,\cS}\left(\psi_{,t}A^*+\psi_{,t}^*A\right) \;,\notag \\
&T^t_{\;x}(\Bx,\tau)=\frac{\tau\left(B+\cS\right)T_{tx}(\Bx)}{2\tilde\tau\,\cS} \;,\quad T^t_{\;\,t}(\Bx,\tau)=\frac{V}{1+\tau V} -\frac{1}{2\tilde\tau\,\cS}\left(B-\cS-\frac{\tilde\tau}{m}\left(A^*\psi_{,x}+A\psi_{,x}^*\right)\right) \;,
\end{align}
obtained from \eqref{eq:stressgen} setting $\cL=\cL(\Bx,\tau)$. This proves that \eqref{eq:lagdef2} is indeed the $\TbT-$deformed non-relativistic Lagrangian density.
\section{Deformed soliton solutions}
The knowledge of the coordinate transformation provide a useful tool to obtain classical solutions to the deformed theory without explicitly solving them. In this section, we concentrate on the NLS theory and derive the $\TbT-$deformation of some particular soliton solutions. Recall that the NLS Lagrangian density is
\begin{equation}
\label{eq:LagNLS}
\cL_{\NLS}(\Bx)=\frac{i}{2}\left(\psi^*\psi_{,t}-\psi\psi_{,t}^*\right)-\frac{\psi_{,x}\psi_{,x}^*}{2m}-V_{\NLS} \;,\quad V_{\NLS}=g|\psi|^4 \;,
\end{equation}
and the Euler-Lagrange equations associated to it are
\begin{equation}
\label{eq:NLSEoMs}
\begin{cases}
-i\psi_{,t}=\frac{1}{2m}\psi_{,xx}-2g|\psi|^2\psi \\
i\psi_{,t}^*=\frac{1}{2m}\psi_{,xx}^*-2g|\psi|^2\psi^*
\end{cases} \;.
\end{equation}
Starting from a given solution $\psi_0(\Bx)$ to \eqref{eq:NLSEoMs}, we can obtain the corresponding $\TbT-$deformed solution $\psi_0(\Bx,\ta)$ by means of the coordinate transformation, using the strategy described in \cite{Conti:2018tca}. Let us summarise here the main idea of the method. We start from the definition of inverse Jacobian \eqref{eq:J-1}
\begin{equation}
\label{eq:systinvJac}
\begin{cases}
\displaystyle{\frac{\partial t}{\partial t'} = 1+\ta\,T^x_{\;\;x}(\By)}\vspace{0.2cm}\\
\displaystyle{\frac{\partial t}{\partial x'} = -\ta\,T^t_{\;x}(\By)}
\end{cases} \;,\quad 
\begin{cases}
\displaystyle{\frac{\partial x}{\partial t'} = -\ta\,T^x_{\;\;t}(\By)}\vspace{0.2cm}\\
\displaystyle{\frac{\partial x}{\partial x'} = 1+\ta\,T^t_{\;\,t}(\By)}
\end{cases} \;,
\end{equation}
and plug the solution $\psi_0(\By)$ together with its c.c. inside the explicit expressions of the components of $T(\By)$. In this way, we end up with two systems of partial differential equations for the unknown functions $t(\By)$ and $x(\By)$. Integrating these systems we first recover the map $\Bx(\By)=\left(t(\By),x(\By)\right)$ and then we invert it to arrive at $\By(\Bx)=\left(t'(\Bx),x'(\Bx)\right)$. Finally, plugging $\By(\Bx)$ into the explicit expression of $\psi_0(\by)$, we obtain the desired deformed solution as
\begin{equation}
\psi_0(\Bx,\ta) = \psi_0(\By(\Bx)) \;.
\end{equation}
\subsection{The bright soliton}
\label{sec:BS}
Bright solitons are solutions localized in space that emerge in the regime $g<0$. Therefore, throughout this section we shall fix $g=-k$, with $k\in\R^+$. The bright soliton solution has the following analytic expression
\begin{equation}
\label{eq:1sol}
\psi_{0}(\By) = \eta\sech{\left(f(\By)\right)}\exp{\left\lbrace i\left(\kappa x'-\omega t'\right)\right\rbrace}\;,\quad f(\By)=\eta\sqrt{2mk}\left(x'-vt'\right) \;,
\end{equation}
where $\eta\in\R^+$ is the amplitude, $\kappa\in\R$ is the wave number, $\displaystyle{\omega=\frac{\kappa^2}{2m}-k\eta^2}$ is the dispersion relation and $\displaystyle{v=\frac{\partial \omega}{\partial \kappa}=\frac{\kappa}{m}}$ is the velocity of the soliton. It can be easily checked that \eqref{eq:1sol} together with its c.c. are solutions to \eqref{eq:NLSEoMs}. Plugging (\ref{eq:1sol}) and its c.c. in \eqref{eq:systinvJac} we arrive at
\begin{gather}
\begin{cases}
\displaystyle{\frac{\partial t}{\partial t'} = 1-\frac{\kappa^2\eta^2\ta}{m}\sech^2{\left(f(\By)\right)}}\vspace{0.1cm}\\
\displaystyle{\frac{\partial t}{\partial x'} = \kappa\eta^2\ta\sech^2{\left(f(\By)\right)}}
\end{cases} \;,\\
\begin{cases}
\displaystyle{\frac{\partial x}{\partial t'} = -\frac{\kappa\eta^2\ta}{2m}\sech^2{\left(f(\By)\right)}\left(\frac{\kappa^2}{m}-2k\eta^2-4k\eta^2\tanh^2{\left(f(\By)\right)}\right)} \notag\\
\displaystyle{\frac{\partial x}{\partial x'} = 1+\frac{\eta^2\ta}{2}\sech^2{\left(f(\By)\right)}\left(\frac{\kappa^2}{m}+2k\eta^2-4k\eta^2\sech^2{\left(f(\By)\right)}\right)}
\end{cases} \;.
\end{gather}
The latter systems of differential equations can be integrated for $\Bx(\By)=\left(t(\By),x(\By)\right)$ as follows
\begin{align}
\label{eq:xtsol}
t(\By)&=t'+\frac{\kappa\eta\ta}{\sqrt{2mk}}\tanh{\left(f(\By)\right)}\;,\notag\\
x(\By)&=x'+\frac{\eta\ta}{3\sqrt{2mk}}\left(-2k\eta^2\sech^2{\left(f(\By)\right)} -k\eta ^2 +\frac{3 \kappa ^2}{2m}\right)\tanh{\left(f(\By)\right)}\;.
\end{align}
where the constants of integration has been chosen in accordance with the initial condition at $\ta=0$. To obtain the inverse relation $\By(\Bx)=\left(t'(\Bx),x'(\Bx)\right)$, we first observe that
\begin{equation}
\label{eq:psif}
|\psi_{0}(\By)|^2=\eta^2\sech^2{\left(f(\By)\right)} \quad\Longrightarrow\quad f(\By) = \text{arcsech}\left(\frac{|\psi_{0}(\By)|}{\eta}\right) \;.
\end{equation}
Plugging \eqref{eq:psif} into \eqref{eq:xtsol} and using the property
$$\tanh{\left(\text{arcsech}(z)\right)}=\sqrt{1-z^2} \;,\quad -1\leq z\leq 1\;,$$
we get
\begin{align}
t(\By)&=t'+\frac{\kappa\ta}{\sqrt{2mk}}\sqrt{\eta^2-|\psi_{0}(\By)|^2}\notag \;,\\
x(\By)&=x'+\frac{\ta}{3\sqrt{2mk}}\left(\frac{3 \kappa ^2}{2 m}-k\left(\eta^2+2|\psi_{0}(\By)|^2\right)\right)\sqrt{\eta^2-|\psi_{0}(\By)|^2} \;.
\end{align}
Since by construction $\psi_{0}(\Bx,\ta)=\psi_{0}(\By(\Bx))$, we have
\begin{align}
t'(\Bx)&=t-\frac{\kappa\ta}{\sqrt{2mk}}\sqrt{\eta^2-|\psi_{0}(\Bx,\ta)|^2}\notag \;,\\
x'(\Bx)&=x-\frac{\ta}{3\sqrt{2mk}}\left(\frac{3 \kappa ^2}{2 m}-k\left(\eta^2+2|\psi_{0}(\Bx,\ta)|^2\right)\right)\sqrt{\eta^2-|\psi_{0}(\Bx,\ta)|^2} \;.
\end{align}
\begin{figure}[t]
  \centering
  \subfloat[]{\includegraphics[scale=0.27]{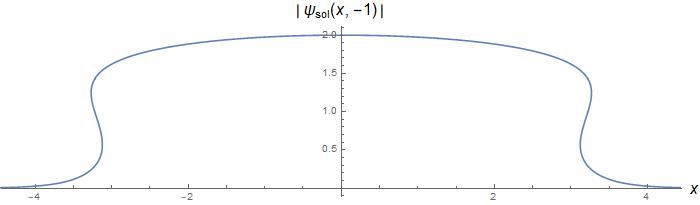}
  \label{fig:sol_tm1}}
  \hspace{0.5cm}
  \subfloat[]{\includegraphics[scale=0.27]{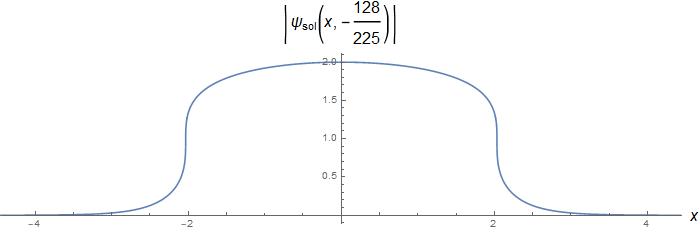}
  \label{fig:sol_tm128over225}} 
  \\
  \subfloat[]{\includegraphics[scale=0.27]{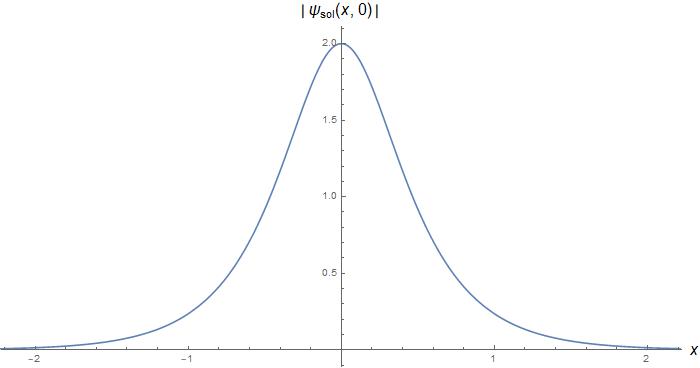}
  \label{fig:sol_t0}}
  \\
  \subfloat[]{\includegraphics[scale=0.27]{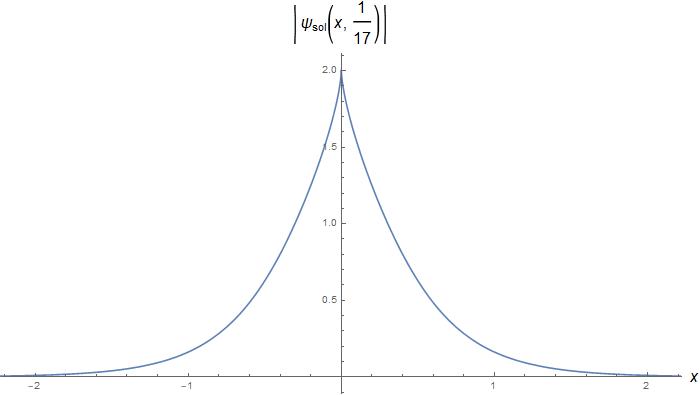}
  \label{fig:sol_tp1over17}}
  \hspace{0.5cm}
  \subfloat[]{\includegraphics[scale=0.27]{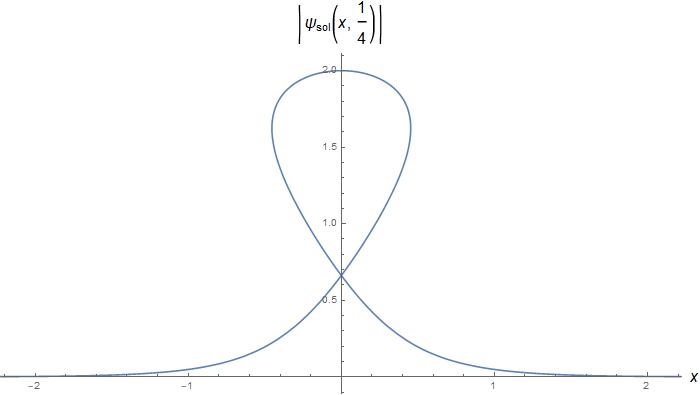}
  \label{fig:sol_tp1over4}}
  \vspace{0.1cm}
\caption{The modulus of the $\textsc{T} \bar{\textsc{T}}$-deformed bright soliton solution $\psi_{0}(\Bx,\ta)$ at $t=0$ and for different values of $\ta$. The parameters are chosen as follows: $\kappa=-1/\sqrt{2}$, $\eta=2$, $m=1$ and $k=1$. Since $\omega=-15/4<0$, the critical values of $\ta$ are $\ta_{crit}^+=1/17$ and $\ta_{crit}^-=-128/225$, according to \eqref{eq:tacrit}.}
  \label{fig:sol}
\end{figure}
In conclusion, $|\psi_{0}(\Bx,\ta)|$ is defined through the following implicit relation
\begin{multline}
f(\Bx) =\text{arcsech}\left(\frac{|\psi_{0}(\Bx,\ta)|}{\eta}\right)- \frac{\eta\ta}{3}\left(\frac{3 \kappa ^2}{2 m}+k\left(\eta^2+2|\psi_{0}(\Bx,\ta)|^2\right)\right)\sqrt{\eta^2-|\psi_{0}(\Bx,\ta)|^2}\;.
\end{multline}
Driven by the analogy with the relativistic case, we expect that the deformation causes the emergence of shock-wave singularities in the solution for some specific critical values of $\ta$, in correspondence of which the solution becomes multi-valued. For these values of $\ta$ the coordinate transformation is not invertible anymore, hence they are defined by the locus
\begin{equation}
\label{eq:detlocus}
\det{\left(J^{-1}(\By)\right)}|_{\psi(\By)=\psi_0(\By)}=0 \;.
\end{equation}
The explicit computation of the determinant of $J^{-1}(\By)$ evaluated on the solution $\psi_0(\By)$ gives
\begin{multline}
\det{\left(J^{-1}(\By)\right)}|_{\psi(\By)=\psi_0(\By)}=\left(\left(1-\frac{\eta ^2 \kappa ^2}{2 m}\ta+k\eta^4\ta\right)\cosh^2{\left(f(\By)\right)}
+\frac{1}{8}\cosh{\left(4f(\By)\right)}-\frac{1}{8}-2k\eta^4\ta\right)\\
\times\sech^4{\left(f(\By)\right)} = 1-\ta|\psi_{0}(\By)|^2\left(\omega+2k|\psi_{0}(\By)|^2\right)\;,
\end{multline}
where in the last equality we used \eqref{eq:psif}. Since $0\leq|\psi_{0}(\By)|\leq \eta$, the values of $\ta$ that fulfil \eqref{eq:detlocus} is given by the image of the real-valued function
\begin{equation}
\label{eq:detequal0}
F(z)=\frac{1}{z^2\left(\omega+2kz^2\right)}\;,\quad 0\leq z\leq \eta \;,
\end{equation}
that is obtained by solving the equation $\det{\left(J^{-1}(\By)\right)}=0$ w.r.t. $\ta$. Recall that the parameter $\omega$ is defined as $\displaystyle{\omega=\frac{\kappa^2}{2m}-k\eta^2}$ with $m,k,\eta\in\R^+$ and $\kappa\in\R$. An elementary analysis of the function $F$ reveals that its domain and image are
$$\text{dom}(F)=\begin{cases}
]0;\eta] \;,\quad \omega>0 \\
\displaystyle{]0;\eta]-\left\lbrace\sqrt{-\frac{\omega}{2k}}\right\rbrace} \;,\quad \omega<0
\end{cases}\;,\quad \text{Im}(F) = \begin{cases}
[\ta_{crit}^+;+\infty] ,\quad \omega>0 \\
]-\infty;\ta_{crit}^-]\cup[\ta_{crit}^+;+\infty[ \;,\quad \omega<0
\end{cases}$$
where
\begin{equation}
\label{eq:tacrit}
\ta_{crit}^+=\frac{1}{\eta^2\left(\omega+2k\eta^2\right)}\;,\quad \ta_{crit}^-=-\frac{8k}{\omega^2} \;.    
\end{equation}
Therefore, the shock-wave phenomenon occurs only for positive $\ta$ when $\omega>0$, and for both positive and negative sign of $\ta$ when $\omega<0$. The latter situation is similar to the sine-Gordon breather \cite{Conti:2018tca}.\\
It is worth notice that there is a strong resemblance between figures \ref{fig:sol} and the plots in \cite{hasimoto_1972}, where the shape and time evolution of vortex filaments are associated with soliton solutions of NLS. This observation suggests that probably the correct framework for the interpretation of the $\TbT$-deformed NLS is through an embedding in three space dimensions, as explained in \cite{hasimoto_1972, rogers2002backlund}. We expect that the only effect of the $\TbT$ is a deformation of the shape of the filament, without changing the ``solitonic surface''.
This fact corresponds to the non-relativistic analogue of what was previously observed in \cite{Cavaglia:2016oda} and \cite{Conti:2018tca} for deformed massless free bosons and the sine-Gordon model, respectively.
\subsection{The grey soliton}
\label{sec:GS}
Another typical solution of the NLS equation is the grey soliton (see, for example \cite{pit2003bose})  that exists in the regime $g>0$. It has the following analytical expression
\begin{equation}
\psi_{0}(\By) = \sqrt{n_0}\left(i\frac{v}{v_1}+\sqrt{1-\frac{v^2}{v_1^2}}\tanh{\left(f(\By)\right)}\right)e^{-i\mu t} \;,\quad f(\By)=\frac{x'-vt'}{\sqrt{2}\xi}\sqrt{1-\frac{v^2}{v_1^2}} \:,
\end{equation}
where, in the cold-atom framework, $v\in\R$ is the velocity of the soliton, $n_0\in\R^+$ is the ground-state density of condensed atoms, $\mu=2 g n_0\in\R^+$ is the chemical potential, $v_1=\sqrt{\mu/m}\in\R^+$ is the velocity of the first sound and $\xi=1/\sqrt{2m\mu}\in\R^+$ is the healing length. We shall follow the same steps of the previous section. Plugging the solution $\psi_{0}(\By)$ together with its c.c. in \eqref{eq:systinvJac} we get
\begin{gather}
\begin{cases}
\displaystyle{\frac{\partial t}{\partial t'} = 1-gn_0^2\ta+\frac{mv^2\ta}{2g}(\mu-mv^2)}\sech^2{\left(f(\By)\right)}\\
\displaystyle{\frac{\partial t}{\partial x'} = \frac{mv\ta}{2g}(mv^2-\mu)}\sech^2{\left(f(\By)\right)}
\end{cases} \;, \notag\\
\begin{cases}
\displaystyle{\frac{\partial x}{\partial t'} = \frac{v\ta}{2g}(\mu-mv^2)\left(mv^2+\mu\sinh^2{\left(f(\By)\right)}\right)\sech^4{\left(f(\By)\right)}}\\
\displaystyle{\frac{\partial x}{\partial x'} = 1+gn_0^2\ta-\frac{\ta}{2g}(\mu-mv^2)\left(mv^2+\mu\sinh^2{\left(f(\By)\right)}\right)\sech^4{\left(f(\By)\right)}}
\end{cases} \;.
\label{eq:JJ1}
\end{gather}
The systems (\ref{eq:JJ1}) can be integrated for $\Bx(\By)=\left(t(\By),x(\By)\right)$ as follows
\begin{align}
t(\By) &= (1-gn_0\ta)t'-\frac{v\ta}{2g}\sqrt{m\mu-m^2v^2}\tanh{\left(f(\By)\right)} \;,\notag \\
x(\By) &= (1+gn_0\ta)x'+\frac{\ta}{6gm}\sqrt{m\mu-m^2v^2}\left(-\mu-2mv^2+(\mu-mv^2)\sech^2{\left(f(\By)\right)}\right)\notag\\
&\times\tanh{\left(f(\By)\right)}\;.
\end{align}
Using the fact that
\begin{equation}
\label{eq:psif2}
|\psi_{0}(\By)|^2 = n_0+\frac{1}{2g}\left(mv^2-\mu\right)\sech^2{\left(f(\By)\right)} \quad\Longrightarrow\quad f(\By) = \text{arcsech}\left(\frac{2g|\psi_{0}(\By)|^2-\mu}{mv^2-\mu}\right) \;,
\end{equation}
we get
\begin{align}
t(\By) &= \left(1-gn_0^2\ta\right)t'-\frac{mv^2\ta}{2g}\sqrt{-1+\frac{2g}{mv^2}|\psi_{0}(\By)|^2}\;,\notag \\
x(\By) &= \left(1+gn_0^2\ta\right)x'-\frac{v\ta}{3g}\left(mv^2+g|\psi_{0}(\By)|^2\right)\sqrt{-1+\frac{2g}{mv^2}|\psi_{0}(\By)|^2} \;.
\end{align}
In conclusion, the inverse relation $\By(\Bx)=\left(t'(\Bx),x'(\Bx)\right)$ yields
\begin{align}
t'(\Bx) &=\frac{1}{1-gn_0^2\ta}\left(t+\frac{mv^2\ta}{2g}\sqrt{-1+\frac{2g}{mv^2}|\psi_{0}(\Bx,\ta)|^2}\right) \;,\notag \\
x'(\Bx) &=\frac{1}{1+gn_0^2\ta}\left(x+\frac{mv^3\ta}{3g}\left(1+\frac{g}{mv^2}|\psi_{0}(\Bx,\ta)|^2\right)\sqrt{-1+\frac{2g}{mv^2}|\psi_{0}(\Bx,\ta)|^2}\right) \;.
\end{align}
Thus, also in this case the deformed solution is defined through an implicit relation as
\begin{multline}
\frac{1}{2\sqrt{\xi}}\sqrt{1-\frac{v^2}{v_1^2}}\left(\frac{x}{1+gn_0^2\ta}-\frac{vt}{1-gn_0^2\ta}\right) = \text{arcsech}\left(\frac{2g|\psi_{0}(\Bx,\ta)|^2-\mu}{mv^2-\mu}\right)
-\frac{1}{2\sqrt{\xi}}\sqrt{1-\frac{v^2}{v_1^2}}\\
\times \frac{mv^3\ta}{g}\sqrt{-1+\frac{2g}{mv^2}|\psi_{0}(\Bx,\ta)|^2}\left(\frac{mv^2+g|\psi_{0}(\Bx,\ta)|^2}{3mv^2\left(1+gn_0^2\ta\right)}-\frac{1}{2\left(1-gn_0^2\ta\right)}\right) \;.
\end{multline}
Following the same logic adopted in the previous section for the bright soliton, it is possible to find the critical values of $\ta$ where the solution becomes multi-valued.  However, the computation and explicit outcomes are quite involved and not particularly enlightening, thus we decided to omit them.
\begin{figure}[H]
  \centering
  \subfloat[]{\includegraphics[scale=0.35]{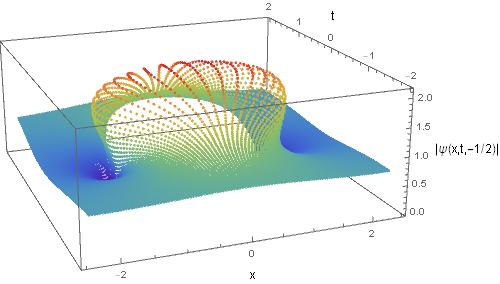}
  \label{fig:per_tm1over2}}
  \hspace{0.5cm}
  \subfloat[]{\includegraphics[scale=0.35]{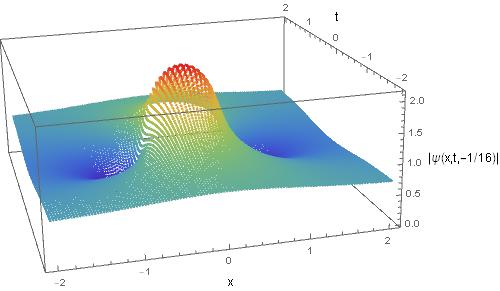}
  \label{fig:per_tm1over8}} 
  \\
  \subfloat[]{\includegraphics[scale=0.35]{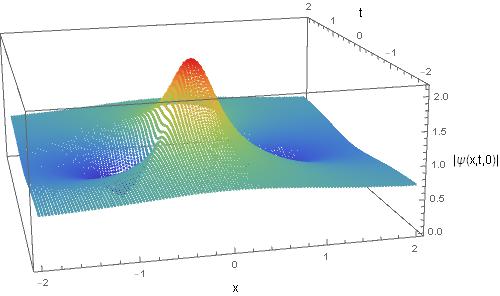}
  \label{fig:per_t0}}
  \\
  \subfloat[]{\includegraphics[scale=0.35]{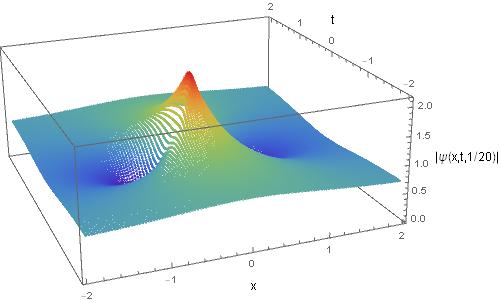}
  \label{fig:per_tp1over20}}
  \hspace{0.5cm}
  \subfloat[]{\includegraphics[scale=0.35]{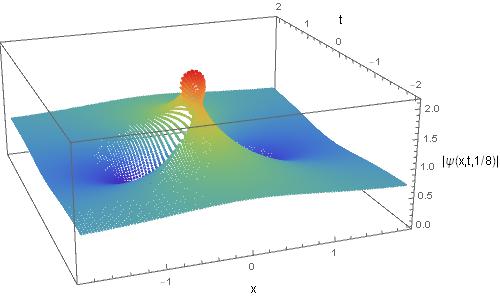}
  \label{fig:per_tp1over8}}
  \vspace{0.1cm}
\caption{The modulus of the $\textsc{T} \bar{\textsc{T}}$-deformed Peregrine's soliton solution $\psi_{0}(\Bx,\ta)$ for different values of $\ta$. The parameters are chosen as follows: $m=1$ and $k=1$.}
  \label{fig:sol1}
\end{figure}
\subsection{The Peregrine's soliton}
As in the case of the bright soliton, we shall set $g=-k$ with $k\in\R^+$. The Peregrine's soliton has the following analytical expression
\begin{equation}
\psi_{0}(\By) = \frac{1}{\sqrt{2k}}\left(1-\frac{4(1+2it')}{1+4mx'^2+4t'^2}\right)e^{it'} \;.
\end{equation}
Plugging the solution and its c.c. in \eqref{eq:systinvJac} we get
\begin{equation}
\begin{cases}
\displaystyle{\frac{\partial t}{\partial t'} = 1-\ta\left(\frac{4\left(1-4t'^2+4mx'^2\right)}{k\left(1+4t'^2+4mx'^2\right)^2}-\frac{1}{4k}\right)}\\
\displaystyle{\frac{\partial t}{\partial x'} = \frac{32m\ta \,t'x'}{k\left(1+4t'^2+4mx'^2\right)^2}}
\end{cases} \;,
\end{equation}
\begin{equation}
\begin{cases}
\displaystyle{\frac{\partial x}{\partial t'} = +32\ta\,t'x'\frac{16t'^4+8t'^2\left(5+4mx'^2\right)+\left(3-4mx'^2\right)^2}{k\left(1+4t'^2+4mx'^2\right)^4}}\\
\displaystyle{\frac{\partial x}{\partial x'} = 1-\ta\frac{\left(16 t'^4+8t'^2 \left(5+4 m x'^2\right)+\left(3-4 m
   x'^2\right)^2\right)^2-1024 mx'^2\left(1+4 t'^2\right)
   }{4k\left(1+4t'^2+4mx'^2\right)^4}}
\end{cases} \;.
\end{equation}
Integrating the latter systems for $\Bx(\By)=\left(t(\By),x(\By)\right)$ we get
\begin{align}
t(\By) &= t'-\frac{\ta\,t'}{4k}\left(-1+\frac{16}{1+4t'^2+4mx'^2}\right)\notag \;,\\
x(\By) &= x'-\frac{\ta\,x'}{12k}\left(3+\frac{256\left(1+4t'^2\right)}{\left(1+4t'^2+4mx'^2\right)^3}-\frac{64}{\left(1+4t'^2+4mx'^2\right)^2}+\frac{48}{1+4t'^2+4mx'^2}\right) \;.
\end{align}
Unfortunately, it is not possible to invert the coordinate transformation, therefore, we resort to numerical integration. Figure \ref{fig:sol1} shows the deformation of the Peregrine soliton for different values of $\tau$. As for the bright and grey solitons, described in sections \ref{sec:BS} and \ref{sec:GS}, we see the appearance of the ``wave-breaking'' phenomena,  resembling that observed in relativistic models \cite{Conti:2018jho, Conti:2018tca, Nastase:2020evb}.
\section{Non-relativistic limit of $\TbT$-deformed sinh-Gordon}
\label{sec:NRL}
As extensively discussed in \cite{Beg:1984yh, Jia:2004qt, Kormos:2009yp, Bastianello_2016}, the non-relativistic limit (NR) of the $\phi^4$ or the sinh-Gordon (sh-G) models correspond to the NLS theory with quartic potential. This limit can be consistently performed not only at the level of the classical action and equations of motion,  but also for various quantum objects of physical interests, such as the Thermodynamic Bethe Ansatz and the form factors.

Therefore, the naive expectation is that the $\TbT$-deformed NLS theory should be easily obtainable from the NR limit of the deformed sinh-Gordon model. We consider the sinh-Gordon theory with background metric $\eta=\text{diag}\left(c^2,-1\right)$ and action
\begin{equation}
\cA_{\shG}=\int dt\,dx\,\sqrt{|\det{\eta}|}\,\cL_{\shG}(\Bx) = \int c\,dt\,dx\,\cL_{\shG}(\Bx) \;,
\end{equation}
where the Lagrangian density is
\begin{gather}
\label{eq:sGLag}
\mathcal{L}_{\shG}(\Bx)=\frac{1}{2}\eta^{\mu\nu}\partial_\mu\phi\,\partial_\nu\phi-V_{\shG} = \frac{1}{2}\left(\frac{\phi_{,t}^2}{c^2}-\phi_{,x}^2\right) - V_{\shG} \;,\quad \mu,\nu\in\{0,1\} \;, \\
V_{\shG}=V=\frac{m^2 c^2}{\bar g^2}\left(\cosh(\bar g\phi)-1\right)\;, 
\end{gather}
with $\phi_{,t}=\partial_0\phi$ and $\phi_{,x}=\partial_1\phi$. The corresponding deformed Lagrangian density is \cite{Conti:2018jho,Bonelli:2018kik}
\begin{equation}
\label{eq:defsGLag}
\mathcal{L}_{\shG}(\Bx,\tau)= -\frac{V}{1-\tau V} + \frac{1}{2\tilde\ta}\left(1 -\sqrt{1- 2\tilde\ta\left(\frac{\phi_{,t}^2}{c^2}-\phi_{,x}^2\right)} \right) \;,\quad \tilde\ta=\ta(1-\ta V) \;,
\end{equation}
that fulfils the flow equation 
\begin{equation}
\label{eq:floweqrel}
\partial_\ta \cL_{\shG}(\Bx,\ta) = c^2\det{\left[T_{\shG}(\Bx,\ta)\right]} \;,
\end{equation}
where the $c^2$ factor comes from $|\det{\eta}|$. In analogy with \cite{Kormos:2009yp}, we shall consider a double scaling limit such that
$$c\to\infty \;,\quad \bar g\to 0 \quad\text{with}\quad \beta = \bar g c = \text{const.} \;.$$
Therefore, the sinh-Gordon potential admits the following expansion around $\bar g=0$
\begin{equation}
V(\phi) = \frac{m^2c^2}{2}\phi^2 +\frac{1}{4!}\beta^2 m^2 \phi^4 +\cO(c^{-2})\;,
\end{equation}
where the powers of $\phi$ higher than $\phi^4$ are suppressed.
Following \cite{Beg:1984yh, Jia:2004qt,Kormos:2009yp}, we  parametrize the field $\phi$  as
\begin{equation}
\label{eq:parametrization}
   \phi(\Bx)=\frac{1}{\sqrt{2m}}\left(e^{imc^2 t}\psi^{*}(\Bx) +e^{-imc^2 t}\psi(\Bx)\right)  \;, 
\end{equation}
where $\psi$ describes only the non-relativistic degrees of freedom. Using \eqref{eq:parametrization}, the kinetic and the potential terms of the sinh-Gordon Lagrangian become
\begin{equation}
V = \frac{mc^2}{2}|\psi|^2 + \frac{\beta^2}{16}
|\psi|^4 + \sum_{n \in \{\pm 2,\pm 4\}} O_n(\Bx)\,e^{inmc^2t} + \cO(c^{-2})\;, 
\label{Vex}
\end{equation}
\begin{eqnarray}
\frac{\phi_{,t}^2}{c^2}-\phi_{,x}^2 &=& -\frac{\psi^{*}_{,x} \psi_{,x}}{m} + i(\psi^{*}\psi_{,t} -\psi\psi^{*}_{,t}) + mc^2|\psi|^2 +  \sum_{n \in \{\pm 2 \} } O'_n(\Bx)\, e^{inmc^2t}+ \cO(c^{-2}) \nonumber \\
&=&2\mathcal{L}_K(\Bx) + m c^2|\psi|^2 + \sum_{n \in \{ \pm 2 \} }O'_n(\Bx) \, e^{inmc^2t}+ \cO(c^{-2})\;, 
\label{eq:phiphiex}
\end{eqnarray}
where $O_n(\Bx)$ and $O'_n(\Bx)$ collect products of powers of $\psi$ and $\psi^{*}$ while $\cL_K(\Bx)$ is the kinetic part of the non-relativistic Lagrangian density $\cL(\Bx)$, as per \eqref{eq:undefLag}. Notice that terms involving exponential factors $e^{inmc^2 t}$ oscillate so fast as $c\to\infty$ that average to zero when integrated over any small but finite time interval. We shall drop these terms taking a suitable time average denoted by the symbol $\langle\star\rangle$. It follows that
\begin{equation}
\langle V\rangle =\frac{mc^2}{2}|\psi|^2 + \frac{\beta^2}{16}
|\psi|^4 + \cO(c^{-2})\;,\quad \left\langle\frac{\phi_{,t}^2}{c^2}-\phi_{,x}^2\right\rangle = 2\mathcal{L}_K(\Bx) + m c^2|\psi|^2 + \cO(c^{-2})\;,
\end{equation}
Plugging (\ref{Vex}) and \eqref{eq:phiphiex} in \eqref{eq:sGLag} and taking the time average, we obtain the result of \cite{Kormos:2009yp}
\begin{equation}
\langle\cL_{\shG}(\Bx)\rangle = \frac{1}{2}\left\langle\frac{\phi_{,t}^2}{c^2}-\phi_{,x}^2\right\rangle -\langle V\rangle =  \cL_{\NLS}(\Bx) + \cO(c^{-2}) \underset{c\to\infty}{\longrightarrow} \cL_{\NLS}(\Bx) \;,
\end{equation}
where $\cL_{\NLS}(\Bx)$ is the nonlinear Schr\"odinger Lagrangian density \eqref{eq:LagNLS} with coupling constant $g=\beta^2/16$, explicitly
\begin{equation}
\label{eq:NLSbeta}
\cL_{\NLS}(\Bx)=\frac{i}{2}\left(\psi^*\psi_{,t}-\psi\psi_{,t}^*\right)-\frac{\psi_{,x}\psi_{,x}^*}{2m}-\frac{\beta^2}{16}|\psi|^4 \;.
\end{equation}
Such non-relativistic limit of the sinh-Gordon model is uniquely defined. However, the same procedure appears to be ambiguous when applied to the $\TbT-$deformed case. 
\subsection{Mean Field approach}
In this section we shall discuss one among the many possible ways to perform the NR limit in the deformed sinh-Gordon model. Before we begin, notice that the factor $c^2$ in \eqref{eq:floweqrel} is problematic when taking the NR limit. Therefore, we reabsorb it by rescaling $\ta$ as $\ta/c^2$ in \eqref{eq:defsGLag}. Given this, the idea is to apply a Mean Field (MF) approach which consists in taking the average of the potential and kinetic terms appearing in \eqref{eq:defsGLag} as follows
\begin{equation}
\label{eq:deflagMF}
\langle\cL_{\shG}(\Bx,\ta)\rangle_{\MF} = -\frac{\langle V\rangle}{1-\frac{\ta}{c^2}\langle V\rangle} + \frac{1}{\frac{2\ta}{c^2}\left(1-\frac{\ta}{c^2}\langle V\rangle\right)}\left(1 -\sqrt{1- \frac{2\ta}{c^2}\left(1-\frac{\ta}{c^2}\langle V\rangle\right)\left\langle\frac{\phi_{,t}^2}{c^2}-\phi_{,x}^2\right\rangle}\right) \;.
\end{equation}
then take the limit $c\to\infty$. This procedure is somehow  justified, a-posteriori,  by the simplicity of the final outcome. Let us consider separately the various terms in expression \eqref{eq:deflagMF}, namely
\begin{equation}
\label{eq:pot1}
-\frac{\langle V\rangle}{1-\frac{\ta}{c^2}\langle V\rangle} = -c^2\frac{\frac{m}{2}|\psi|^2}{1-\ta\frac{m}{2}|\psi|^2} - \frac{\frac{\beta^2}{16}|\psi|^4}{\left(1-\ta\frac{m}{2}|\psi|^2\right)^2} +\mathcal{O}(c^{-1}) \;,
\end{equation}
and
\begin{multline}
\label{eq:sqrt1}
\frac{1}{\frac{2\ta}{c^2}\left(1-\frac{\ta}{c^2}\langle V\rangle\right)}\left(1 -\sqrt{1- \frac{2\ta}{c^2}\left(1-\frac{\ta}{c^2}\langle V\rangle\right)\left\langle\frac{\phi_{,t}^2}{c^2}-\phi_{,x}^2\right\rangle}\right) \\
=c^2\frac{\frac{m}{2}|\psi|^2}{1-\ta \frac{m}{2}|\psi|^2} +\frac{\cL_{\NLS}(\Bx)}{1-\ta m|\psi|^2} 
+\frac{\frac{\beta^2}{16}|\psi|^4}{\left(1-\ta\frac{m}{2}|\psi|^2\right)^2} + \mathcal{O}(c^{-1}) \;,
\end{multline}
where $\mathcal{L}_{\NLS}(\Bx)$ is as per \eqref{eq:NLSbeta}. Combining \eqref{eq:pot1} and \eqref{eq:sqrt1}, the terms proportional to $c^2$ cancel in the sum and lead to
\begin{equation}
\label{eq:Lt}
\langle\cL_{\shG}(\Bx,\ta)\rangle_{\MF} = \frac{\cL_{\NLS}(\Bx)}{1-\ta m|\psi|^2} + \mathcal{O}(c^{-1}) \underset{c\to\infty}{\longrightarrow} \frac{\cL_{\NLS}(\Bx)}{1-\ta m|\psi|^2}=\cL_{\NR}(\Bx,\ta) \;.
\end{equation}
This is an unexpected result, since \eqref{eq:Lt} is very different from  the $\TbT$-perturbed Lagrangian derived in section \ref{DeformedL}. Although the question is certainly open, we think it is unlikely that the Lagrangian  $\mathcal{L}_{\NR}(\Bx,\tau)$ corresponds to an integrable deformation of NLS.

The corresponding flow equation is
\begin{equation}
\label{op}
\partial_\tau \mathcal{L}_{\NR}(\Bx,\tau)=\frac{m|\psi|^2}{1-\tau m|\psi|^2} \mathcal{L}_{\NR}(\Bx,\tau) \;.
\end{equation}
The Hamiltonian density is
\begin{equation}
\label{eq:HNR}
\mathcal{H}_{\NR}(\Bx,\tau)=\pi  \psi_{,t}  + \pi^{*}\psi^{*}_{,t} - \mathcal{L}_{\NR}(\Bx,\tau) = \frac{\cH_{\NLS}(\Bx)}{1-\ta m|\psi|^2}  \;,
\end{equation}
where $\displaystyle{\cH_{\NLS}(\Bx)=\frac{\psi_{,x}\psi_{,x}^*}{2m}+\frac{\beta^2}{16}|\psi|^4}$ and the conjugated momenta are
\begin{equation}\label{mom}
\pi=\frac{\partial \mathcal{L}_{\NR}(\Bx,\tau)}{\partial \psi_{,t}}= \frac{\frac{i}{2}\psi^{*}}{1-\ta m|\psi|^2} \;,\quad \pi^{*}=\frac{\partial \mathcal{L}_{\NR}(\Bx,\tau)}{\partial \psi^{*}_{,t}}= \frac{-\frac{i}{2}\psi}{1-\ta m|\psi|^2}\;.
\end{equation}
It is easy to show that all components of the stress-energy tensor rescale in the same way:
\begin{equation}
T_{\NR}(\Bx,\tau)=\frac{T_{\NLS}(\Bx)}{1-\ta m|\psi|^2}\;,
\end{equation}
where $T_{\NLS}(\Bx)$ is as per \eqref{eq:TNLS} with $\cL=\cL_{\NLS}$. As it is customary in the integrable model framework, it is convenient to continue working with the field pair $(\psi,\psi^*)$. The equal-time Poisson bracket of the deformed theory is
\begin{equation}
\label{eq:PoissonNR}
\boxed {\{ \psi(x), \psi^*(y) \} = -i\left(1-\ta m\,\psi(x)\,\psi^*(x)\right)^2 \, \delta(x-y)\;.} 
\end{equation}
where $x$ and $y$ denote two different spatial points at fixed time and $\delta(x)$ is the Dirac delta. In fact, using the former definition it is possible to verify that formula
\begin{equation}
\psi^*_{,t}(x) = \{\psi^*(x),H_{\NR}(\tau)\}\;,\quad H_{\NR}(\ta)= \int dx \;\cH_{\NR}(\Bx,\ta) \;,
\end{equation}
together with its c.c. yield the deformed EoMs. Let us briefly sketch the computation. From the definition \eqref{eq:HNR} and using Leibnitz rule we have
\begin{multline}
\label{eq:poisson1}
\{\psi^*(x),H_{\NR}(\tau)\} = \int dy\,\left\lbrace\psi^*(x),\frac{\cH_{\NLS}(y)}{1-\ta m\,\psi(y)\,\psi^*(y)}\right\rbrace \\
=\int dy\,\frac{1}{1-\ta m\,\psi(y)\,\psi^*(y)}\lbrace\psi^*(x),\cH_{\NLS}(y)\rbrace + \cH_{\NLS}(y)\left\lbrace\psi^*(x),\frac{1}{1-\ta m\,\psi(y)\,\psi^*(y)}\right\rbrace \;.
\end{multline}
Next, we manipulate separately the Poisson brackets in the second line of the latter expression, obtaining
\begin{multline}
\label{eq:firstterm}
\lbrace\psi^*(x),\cH_{\NLS}(y)\rbrace=\frac{\psi_{,y}^*(y)}{2m}\lbrace\psi^*(x),\psi_{,y}(y)\rbrace + 2g\,\psi(y)\left(\psi^*(y)\right)^2\lbrace\psi^*(x),\psi(y)\rbrace \\
=\psi^*_{,y}(y)\left(1-\ta m\,\psi(y)\,\psi^*(y)\right)\Bigl(\frac{i}{2m}\left(1-\ta m\,\psi(y)\,\psi^*(y)\right)\delta_{,y}(x-y) -i\ta\,\partial_y\left(\psi(y)\,\psi^*(y)\right)\delta(x-y)\\+2ig\,\psi(y)\,\psi^*(y)\left(1-\ta m\,\psi(y)\,\psi^*(y)\right)\delta(x-y)\Bigr)\;,
\end{multline}
and
\begin{eqnarray}
\label{eq:secondterm}
\left\lbrace\psi^*(x),\frac{1}{1-\ta m\,\psi(y)\,\psi^*(y)}\right\rbrace &=& \left\lbrace\psi^*(x),\sum_{n\geq 0}\left(\ta m\,\psi(y)\,\psi^*(y)\right)^n\right\rbrace \nonumber\\
= \sum_{n\geq 0}\left(\ta m\right)^n\left\lbrace \psi^*(x),\left(\psi(y)\,\psi^*(y)\right)^n\right\rbrace &=& \frac{1}{\psi(y)}\lbrace\psi^*(x),\psi(y)\rbrace\sum_{n\geq 0}n\left(\ta m\,\psi(y)\,\psi^*(y)\right)^n \nonumber \\
&=&i\ta m\,\psi^*(y)\,\delta(x-y) \;,
\end{eqnarray}
where we used the fact that
\begin{eqnarray}
\lbrace\psi_{,x}(x),\psi^*(y)\rbrace &=&-i\left(1-\ta m\,\psi(x)\,\psi^*(x)\right)^2\delta_{,x}(x-y) \nonumber \\
&+&2im\ta\,\partial_x\left(\psi(x)\,\psi^*(x)\right)\left(1-\ta m\,\psi(x)\,\psi^*(x)\right)\delta(x-y) \;,
\end{eqnarray}
that follows immediately from \eqref{eq:PoissonNR}. Finally, plugging \eqref{eq:firstterm} and \eqref{eq:secondterm} into \eqref{eq:poisson1} we get
\begin{multline}
\label{eq:poisson2}
\{\psi^*(x),H_{\NR}(\tau)\} = -i\ta\,\psi_{,x}^*(x)\,\partial_x\left(\psi(x)\,\psi^*(x)\right) +2ig\,\psi(x)\left(\psi^*(x)\right)^2\left(1-\ta m\,\psi(x)\,\psi^*(x)\right) \\
+i\ta m\,\psi^*(x)\,\cH_{\NLS}(x)+ \frac{i}{2m}\int dy\,\psi_{,y}^*(y)\left(1-\ta m\,\psi(y)\,\psi^*(y)\right)\delta_{,y}(x-y) \;.
\end{multline}
Integrating by parts the integral in \eqref{eq:poisson2}
\begin{eqnarray}
\frac{i}{2m}\int dy\,\psi_{,y}^*(y)\left(1-\ta m\,\psi(y)\,\psi^*(y)\right)\delta_{,y}(x-y) &=& -\frac{i}{2m}\psi_{,xx}^*\left(1-\ta m\,\psi(x)\,\psi^*(x)\right) \nonumber\\
&+&\frac{i\ta}{2}\,\psi_{,x}^*(x)\,\partial_x\left(\psi(x)\,\psi^*(x)\right) \;,
\end{eqnarray}
we find
\begin{eqnarray}
\psi_{,t}^*=\{\psi^*(x),H_{\NR}(\tau)\} &=& -\frac{i}{2m}\psi_{,xx}^*\left(1-\ta m\,\psi(x)\,\psi^*(x)\right) -\frac{i\ta}{2}\,\psi(x)\left(\psi_{,x}^*(x)\right)^2 \nonumber \\
&+&2ig\,\psi(x)\left(\psi^*(x)\right)^2\left(1-\frac{\ta m}{2}\,\psi(x)\,\psi^*(x)\right) \;,
\end{eqnarray}
that is equivalent to 
\begin{equation}
\boxed{\left(i\psi_{,t}^*-\frac{1}{2m}\psi_{,xx}^*+2g\psi(\psi^*)^2-\ta m\left(g\psi^2(\psi^*)^3+\frac{1}{2m}\psi(\psi_{,x}^*)^2-\frac{1}{2m}|\psi|^2\psi_{,xx}^*\right)\right) = 0 \;.}
\label{eq:EoMcinfy}
\end{equation}
The  same EoM  can be directly derived from  the  Lagrangian \eqref{eq:Lt}, however the knowledge of the Poisson structure \eqref{eq:PoissonNR} should help in the exploration of the hidden integrability  structure and for the direct quantisation of this model. Summing up \eqref{eq:EoMcinfy} with its complex conjugate, we can derive the deformed continuity equation
\begin{equation}
\partial_t\left( \frac{m|\psi|^2}{1- \ta m |\psi|^2}\right) +\partial_x\left( \frac{i}{2}\frac{\psi \psi^{*}_{,x}-\psi^{*} \psi_{,x}}{1- \ta m |\psi|^2}  \right) =0 \;,
\label{eq:cont}
\end{equation}
from which we read off the components of the deformed conserved $U(1)$ current
\begin{equation}
J_t(\Bx,\ta) = \frac{m|\psi|^2}{1- \ta m |\psi|^2} \;,\quad J_x(\Bx,\ta) = \frac{i}{2}\frac{\psi \psi^{*}_{,x}-\psi^{*} \psi_{,x}}{1- \ta m |\psi|^2} \;.
\end{equation}
Splitting the complex function $\psi$ in its modulus and phase as
\begin{equation}
\psi= \sqrt{\rho} e^{i \varphi} \;,\quad  v=\fract{1}{m} \partial_x \varphi \;,
\end{equation}
we can recast equation (\ref{eq:cont}) in the form
\begin{equation}
\label{eq:continuity}
\partial_t \left( \frac{\rho}{1-\ta m \rho}\right) + \partial_x\left( \frac{\rho \, v}{1-\ta m \rho}  \right) =0 \;.
\end{equation}
While we were working  on this project the paper  \cite{Cardy:2020olv} appeared on ArXiv. 
Following a quite different logic, using the ideas of a change of the metric and  generalized hydrodynamics the authors  of \cite{Cardy:2020olv} also arrived to (\ref{eq:continuity}).
At this point, it is clear that the non-relativistic limit's outcome depends on the stage chosen to perform the ``small-time interval'' average procedure. We have not yet identified a guiding principle to discern between the various options. Let us end this section with a concluding remark. If we parametrize the field $\phi$ as per \eqref{eq:parametrization}, employ the notion of time-average as described in the previous section and finally take the $c\to\infty$ limit we obtain the following relation
\begin{equation}
\det{\left[\langle T_{\shG}(\Bx)\rangle\right]} = \frac{1}{2}\e_{\mu\rho}\e_{\nu\sigma}\langle T_{\shG}^{\mu\nu}(\Bx)\rangle\langle T_{\shG}^{\rho\sigma}(\Bx)\rangle \underset{c\to\infty}{\longrightarrow}-\e_{\mu\nu}\cJ^\mu(\Bx)\cP^{\nu}(\Bx) \;,
\label{eq:dettoJP}
\end{equation}
where $\cP^\mu$ and $\cJ^\mu$ are the conserved currents associated to the nonlinear Schr\"odinger Lagrangian density \eqref{eq:NLSbeta}. Thus, the first order truncation of \eqref{eq:defsGLag} becomes, upon rescaling $\ta$ as $\ta/c^2$,
\begin{equation}
\langle\cL_{\shG}(\Bx)\rangle + \ta \det{\[\langle T_{\shG}(\Bx)\rangle\]} + \mathcal{O}(\ta^2) \underset{c\to\infty}{\longrightarrow} \cL_{\NLS}(\Bx) -\ta \e_{\mu\nu}\cJ^\mu(\Bx)\cP^{\nu}(\Bx) + \mathcal{O}(\ta^2) \;.
\end{equation}
Notice that the RHS of the latter formula is the first order truncation of the solution to the flow equation 
\begin{equation}
\label{eq:hardrodflow}
\partial_\ta \cL(\Bx,\ta) = -\e_{\mu\nu}\cJ^\mu(\Bx,\ta)\cP^\nu(\Bx,\ta)\;,
\end{equation}
with initial condition $\cL(\Bx,0)=\cL_{\NLS}(\Bx)$,  that is, the Lagrangian density associated with the hard-rod deformation of the nonlinear Schr\"odinger model, recently studied in \cite{Hansen:2020hrs}. It would be important to understand if there is a scheme to reconstruct the flow equation \eqref{eq:hardrodflow} from the $\TbT$ flow of the sinh-Gordon model.
\section{Conclusions}
\label{sec:conclusions}
The nonlinear Schr\"odinger equation plays a significant role in various physics branches, ranging from classical hydrodynamics, superfluidity, and nonlinear optics.

In this paper we have  identified the $\TbT$-deformed NLS Lagrangian, with generic interacting potential, and  studied particular solutions of the corresponding equations of motion.

Compared to the unperturbed case, the deformed soliton solutions exhibit the phenomena of bifurcations or wave breaking.
Several aspects of this model deserve further study. First of all, we would like to fully develop the Hamiltonian approach, which is made complicated by the presence, already in the undeformed theory, of a second class Hamiltonian constraint. From the fact that the finite volume/temperature spectrum of $\TbT$-like perturbed models fulfils Burgers-type equations \cite{Smirnov:2016lqw, Cavaglia:2016oda, Guica:2017lia, Chakraborty:2018vja}, we know that there must be a way to overcome the technical problems caused by the  highly non-trivial evolution of the Poisson bracket structure under the  $\TbT$-perturbation. Various quantum aspects of this deformation are discussed in the nice recent work \cite{Jiang:2020nnb}.

The second type of deformation, described in section \ref{sec:NRL}, is also interesting. In many respects, it leads to a simpler system compared to the  ``standard'' $\TbT$ perturbation of section \ref{DeformedL}. For both perturbations, it would be essential to investigate the connection with the theory of vortex filaments, as discussed in \cite{hasimoto_1972} (see also \cite{rogers2002backlund}). Further work in this direction  may shed some light on the physical interpretation of these systems and their possible interpretation as non-relativistic variants of the  Nambu-Goto model.  Finally, it is necessary to stress that under specific conditions, the NLS equation represents a model for solitons and rogue waves in hydrodynamics and nonlinear optics \cite{ONORATO201347, bookSpaceSolitons}. Interestingly in these laboratory setups, one can exchange the role between space and time coordinates and even describe stationary optical beams where both $t$ and $x$ correspond to physical space coordinates. This is for example achieved in planar glass wave-guides with Kerr non-linearity \cite{Aitchison:90, bookSpaceSolitons}.
The possibility to build these type of devices gives us some hope for future realisations of ``$\TbT$-optical systems''  related, for example, to the simple EoM (\ref{eq:EoMcinfy}). 

Furthermore, it  would be important to explore possible connections between the results presented here and in \cite{Jiang:2020nnb} and the corresponding deformations of the  two-dimensional Yang-Mills-Higgs model \cite{Gerasimov_2007}, as recently suggested by \cite{Santilli:2020qvd} as a natural generalization of the $\TbT$ and q-deformed Yang-Mills setups of \cite{Conti:2018jho,Santilli:2018xux, Santilli:2020qvd, Gorsky:2020qge}.

Finally, concerning the extension to other non-Lorentz invariant models, it would be nice to study the $\TbT$-deformed KdV equation, the classical Lagrangian, the corresponding soliton solutions, and the link with the ODE/IM correspondence \cite{Dorey:2009xa, Dorey:2019ngq, Masoero:2018rel, Conti:2020zft}. Understanding, at a deeper level, the  $\TbT$-like deformations of quantum spin-chains \cite{Pozsgay:2019ekd,Marchetto:2019yyt},  possibly within the  Quantum Spectral Curve framework of AdS/CFT \cite{Gromov:2014bva, Gromov:2013pga, Marboe:2018ugv, Anselmetti:2015mda,Bombardelli:2018bqz, Cavaglia:2015nta}, is also an unexplored avenue that might lead to unexpected and exciting discoveries.  
\\

\noindent
\noindent{\bf Note added 1:}
While we were already at the writing stage of this paper, we became aware of the work \cite{Hansen:2020hrs} by Dennis Hansen, Yunfeng Jiang and Jiuci Xu, which has some overlap with ours. In particular, on the dynamical change of coordinates and the $\TbT$-deformed Lagrangian described in section \ref{DeformedL}. 

\noindent
\noindent{\bf Note added 2:} We thank Sergey Frolov for informing us that, in collaboration with Chantelle Esper, obtained the one-soliton solution of the $\TbT$-deformed NLS equation \cite{EsperFrolov2021} using the light-cone gauge approach of \cite{Frolov:2019nrr}.

\medskip
\noindent{\bf Acknowledgements --}
We would like to thank Miguel Onorato,  Yunfeng Jiang, Leonardo Santilli and Sergey Frolov for useful discussions. This work was also partially supported by the INFN project SFT and by the FCT Project PTDC/MAT-PUR/30234/2017 ``Irregular connections on algebraic curves and Quantum Field Theory``. R.C. is supported by the FCT Investigator grant IF/00069/2015 ``A mathematical framework for the ODE/IM correspondence''.

\bibliography{Biblio3} 
\end{document}